\documentclass[aps,pra,12pt,groupedaddress,showkeys,superscriptaddress,showpacs,reprint,raggedbottom,onecolumn,longbibliography]{revtex4-1}
\usepackage{amsmath}
\usepackage{amsthm}
\usepackage{bm}
\usepackage{graphicx}
\usepackage{hyperref}
\hypersetup{
    colorlinks
}
\allowdisplaybreaks
\usepackage{wrapfig}
\usepackage{multirow}
\usepackage{nameref}
\usepackage{array}
\begin{document}

\title
{
Development of effective stochastic potential method 
using random matrix theory 
for efficient conformational sampling 
of semiconductor nanoparticles 
at non-zero temperatures
}

\author{Jeremy A. Scher}
\affiliation
{
Department of Chemistry, Syracuse University, Syracuse, New York 13244 USA
}

\author{Michael G. Bayne}
\affiliation
{
Department of Chemistry, Syracuse University, Syracuse, New York 13244 USA
}

\author{Amogh Srihari}
\affiliation
{
Department of Biomedical and Chemical Engineering, Syracuse University, Syracuse, New York 13244 USA
}

\author{Shikha Nangia}
\affiliation
{
Department of Biomedical and Chemical Engineering, Syracuse University, Syracuse, New York 13244 USA
}
\email{snangia@syr.edu}

\author{Arindam Chakraborty}
\affiliation
{
Department of Chemistry, Syracuse University, Syracuse, New York 13244 USA
}
\email{archakra@syr.edu}


%\begin{document}
%\maketitle
%\input{001_sec_abstract_jctc}
%
\begin{abstract} 
The relationship between structure and property is central to chemistry, and enables the understanding of chemical phenomena and processes. Need for an efficient conformational sampling of chemical systems arises from the presence of solvents and the existence of non-zero temperatures. However, conformational sampling of structures to compute molecular quantum mechanical properties is computationally expensive because a large number of electronic structure calculations are required. 
In this work, the development and implementation of the effective stochastic potential (ESP) method is presented to perform efficient conformational sampling of molecules. The overarching goal of this work is to alleviate the computational bottleneck associated with performing a large number of electronic structure calculations required for conformational sampling.  
We introduce the concept of a deformation potential and demonstrate its existence by the proof-by-construction approach. 
A statistical description of the fluctuations in the deformation potential due to non-zero temperature was obtained using infinite-order moment expansion of the distribution.  
The formal mathematical definition of the ESP was derived using functional minimization approach to 
match the infinite-order moment expansion for the deformation potential.  
Practical implementation of the ESP was obtained using the random-matrix theory method. 
The developed method was applied to two proof-of-concept calculations of 
the distribution of HOMO-LUMO gap in water molecule and solvated CdSe clusters at 300K.
The need for large sample size to obtain statistically meaningful results was demonstrated 
by performing $10^5 $ ESP calculations. The results from these prototype calculations demonstrated the efficacy of the ESP method for performing efficient conformational sampling. We envision that the fundamental nature of this work will not only extend our knowledge of chemical systems at non-zero temperatures but will also generate new insights for innovative technological applications. 
\end{abstract} 
\maketitle
\section{Introduction} \label{sec:intro}
The goal of this work is to develop an efficient theoretical and computational method for performing ensemble-averaged properties of semiconductor nanoparticles (NPs). Semiconductor NPs such as quantum dots and rods have important technological applications. Because of the quantum confinement effect~\cite{Wang1991525}, their optoelectronic properties can be tuned by adjusting parameters such as the size, shape, material composition, and ligands of the NP. Consequently, NPs have promising applications in photovoltaics,~ 
\cite{Alivisatos199613226,Zidek20126393,Santra20122508,Tang20124889,Salant20122095} light emitting devices,~\cite{Boehme20132500,Jun20131472,Park2008} 
charge and energy transfer processes~\cite{Huang201216472,Ji20126006,Peng20114562,Zidek201212110}, and biological labeling.~\cite{Zhang20133896,BruchezJr.19982013} 
\par     
One of the important environmental factors that influence their optical and electronic properties is the temperature, which has been the focus of numerous theoretical and experimental studies. Early experimental investigation performed by Joshi and co-workers in 2006 looked at this temperature dependence of band gap energies for colloidal CdSe/ZnS core/shell nanocrystals of different diameters. They found that over a temperature range of 10K to 300K, the band gap energies decreased with increasing temperature for all nanocrystal sizes~\cite{doi:10.1063/1.2357856}. Recent experimental results on the temperature-effect on quantum dot (QD) optical properties have also been obtained by Savchenko and co-workers in 2017. They studied the first exciton peak shift as a function of temperature in InP/ZnS core/shell QDs in the range of 6.5K to 296K. Similarly, to Joshi et al. they found that the peak both broadens and shifts toward larger wavelengths with increasing temperature~\cite{Savchenko:17}. In 2011, Chen and co-workers performed nonadiabatic molecular dynamics simulations on both spherical and elongated CdSe quantum dots using time-domain Kohn-Sham theory. Their results confirm the inverse temperature dependence on band gap energy in quantum dots. They also found that the temperature dependence of the relaxation rate of excited electrons was higher in elongated quantum dots~\cite{olegcdse}. In the same year, Fischer et al. performed time-dependent density functional theory (TD-DFT) calculations on a Si$_{29}$H$_{24}$ quantum dot trajectory. They found that the band gap energy of the QD shifted significantly throughout the dynamics simulation~\cite{doi:10.1063/1.3526297}. More recent quantum dynamics simulations utilizing TD-DFT and the tight-binding approximation performed by High and co-workers looked at the excited state energy transfer in a porphyrin light-harvesting device. Their simulation showed that it is essential to couple the nuclear and electronic degrees of freedom to accurately simulate excited state energy transfer~\cite{doi:10.1021/acs.jpca.6b05739}. Liu and Jakubikova have also investigated the electron transfer rate in a pyridine-TiO$_2$ light harvesting assembly, and found that the transfer rate was faster when sampling from thermally accessible structures at room temperature compared to a single 0K optimized structure~\cite{C7SC01169E}. 
\par      
In colloidal systems, the NPs are surrounded by solvent molecules which provide a constant-temperature heat-bath for the NP. Constant-temperature ensembles such as NVT or NPT are used extensively for calculation and analysis of equilibrium properties. In molecular dynamics simulations, various methods are used to implement the thermostat, such as Berendsen~\cite{doi:10.1063/1.448118}, Langevin~\cite{schlick2010molecular}, and Nos\'{e}-Hoover~\cite{doi:10.1063/1.447334,PhysRevA.31.1695}. Investigation of temperature dependence is complicated and computationally expensive because it requires electronic structure calculations on a large number of deformed structures. This imposes serious limitations on the size of the systems that are being investigated. 
\par 
Disorder arising from temperature or solvent effects can also impact the observed properties of organic molecules and systems as well. Much theoretical and computational work has been performed in this area. In 2001, Kwasniewski and coworkers calculated UV-vis absorption spectra on various frames throughout molecular dynamics simulations of trans-stilbene using the ZINDO-CIS method. They found that thermal motion had a  strong impact on the HOMO to LUMO transition energy. At temperatures less than 400K, the energy levels displayed pronounced Gaussian thermal broadening of the related band, up to 30 nm~\cite{Kwasniewski2001557}. In 2003, Lewis and co-workers performed MD simulations combined with tight-binding density functional theory to calculate the electronic states of a model DNA double helix as the molecule underwent room temperature thermal fluctuations. They found that the population density of the HOMO state over time changed dramatically due to these thermal fluctuations~\cite{Lewis20032581}. In 2006, Morth and Autschbach investigated the temperature dependence of the optical rotation of fenchone using TD-DFT. They found that optical rotation in this system increases linearly with temperature over the range of 273K-373K, and concluded that the temperature-dependent vibrational effects were largely responsible for the temperature-dependent rotational effects in rigid organic molecules~\cite{Mort200611381}. 
 
\par 
Many researchers have also investigated these solvent and temperature effects on organic chemical systems experimentally. In 2002, Ariu et al. studied the temperature dependence on photoluminescence quantum yields (PLQYs) of poly(9, 9-dioctylfluorene) films. They found that the PLQYs increased as the temperature of many of these systems were reduced. In 2005, Dauphas and co-workers investigated $\beta$-casein protein aggregation and properties as a function of concentration, temperature, and calcium content, characterized by dynamic light scattering, confocal microscopy, and fluorescence spectroscopy. Over a temperature range of 283K to 323K, they saw that the fluorescence absorption profiles of the protein were very different when the concentration and temperature were changed. In 2014, Mani et al. looked at the fluorescent properties of the salt of 2,6-diaminopyridinium with dihydrogen phosphate. They found that as a result of proton transfer from phosphoric acid in solution to the pyridine nitrogen, the band gap between the HOMO and LUMO states is smaller than compared to 2,6-diaminopyridine. This finding was confirmed by a red shift that was observed in its absorption spectrum~\cite{Mani20146883}. 
 
\par      
In this work, we present the effective stochastic potential (ESP) method to address the computational bottleneck associated with performing electronic structure calculations on a large number of molecular structures needed for obtaining ensemble-averaged quantities. The central idea of the ESP method is to approximate the explicit interactions between the system and the bath degrees of freedom by a statistically equivalent effective potential. In the ESP method, the construction of the effective potential is performed using the random-matrix theory (RMT) method. 
Before we describe this new method, a brief background in the history and usage of RMT will be provided for context. Originally, RMT was designed by Wigner in 1951 to deal with the statistics of eigenvalues and eigenfunctions of complex many-body quantum nuclear systems~\cite{10.2307/1970079}. Wigner later used RMT to investigate the statistical fluctuations of scattering processes.~\cite{Guhr1998189} 
After Wigner, RMT was successfully applied to the description of spectral fluctuations of atomic nuclei~\cite{Bohigas1992}, complex atoms~\cite{Garg1964}, and complex molecules~\cite{HAUSSER1968329}. The nature of the general theory greatly lends itself to the investigation of highly complex and chaotic systems. 
Other, more recent examples of how RMT has been applied to physical systems include equilibrium and transport properties of disordered and classically chaotic quantum systems~\cite{ericson1966fluctuations}, transitions between classical and quantum distributions of hydrogen energy levels~\cite{PhysRevA351464}, and spectral resonances in quartz~\cite{PhysRevLett774918}. Much work has also been done using RMT to investigate the quantum mechanical phenomenon of so-called "persistent current," the perpetual flow of interacting electrons in a static magnetic field~\cite{Montambaux1996,PhysRevB504943,PhysRevB5210772,02955075375347,02955075296008}. 
\par There are many forms of RMT that have been used, but in general, RMT is the process of replacing the Hamiltonian of a system with an ensemble of Hamiltonians containing random matrix elements. Most popularly used in RMT is the Gaussian orthogonal ensemble (GOE), where the Hamiltonian is partitioned into two parts: a reference component and a stochastic component where the matrix elements are uncorrelated, normally distributed random numbers. These two components are then coupled to form the total Hamiltonian.  
Due to their highly complex nature, QDs are a natural candidate for investigation using RMT.  
One of the earliest investigations of QDs using RMT was done by Jalabert, Stone, and Alhassid in 1992. They developed a statistical theory of the effect of Coulomb blockade in QDs represented as a particle in a finite potential well using random matrices to determine the Coulomb blockade amplitudes. They were able to obtain good agreement with experiment.~\cite{Jalabert19923468} 
In 1996, \u{S}eba et al. used a random matrix model to describe the conductance of a model QD connected to an ideal lead. To accurately match experimental results of QDs attached to two leads, they introduced a coupling constant to their RMT Hamiltonian.~\cite{Šeba199613024} 
In that same year, Sivan and co-workers performed an experimental investigation into the disorder of ground state energies of populations of GaAs QDs.~\cite{Sivan19961123}. 
More recently, in 2008 Kaplan et al. investigated the fluctuation of two-body screened Coulomb interaction in ballistic QDs using the random wave model (a model related to RMT). They were able to derive analytical expressions for these fluctuations in terms of the linear size of the QD. Despite the Gaussian random single electron basis used in their theory, they found that both two-body and one-body matrix elements followed non-Gaussian distributions.~\cite{Kaplan2008} 
Shankar in 2006, and later in 2008, has discussed a technique for studying QDs using the renormalization group method in conjunction with RMT.~\cite{Shankar20061173,Shankar2008379}
In 2017, Akimov utilized stochastic Hamiltonians to investigate excited state relaxation in nonadiabatic condensed matter systems~\cite{10.1021/acs.jpclett.7b02185}.
\par  
The remaining part of this article is organized as follows.  
The main theoretical development of the ESP method is described in \autoref{sec:theory}. 
The derivation includes the definition of the quantum mechanical Hamiltonian (\autoref{sec:def}) 
under investigation, the definition of the deformation potential (\autoref{sec:vdef}), and the conceptual construction of the effective stochastic potential (\autoref{sec:esp}). The computational details for implementation of  
the ESP method are presented in \autoref{sec:implementation}. Specifically, implementation details on the choice of sampling schemes(\autoref{sec:md}), choice of one-particle basis functions (\autoref{sec:basis}), and stochastic sampling using Gaussian random matrices (\autoref{sec:gauss}) are presented . The developed ESP method was used to perform proof-of-concept calculations of ensemble-averaged HOMO-LUMO gap energies in chemical systems, and results for $\mathrm{H}_2 \mathrm{O}$ and $\mathrm{Cd}_{20}\mathrm{Se}_{19}$ are presented in \autoref{sec:results} and the discussion of the results are presented in \autoref{sec:discuss}. 
\section{Theory} \label{sec:theory}
\subsection{System setup and definitions} 
\label{sec:def} 
We start by introducing key concepts and definitions  
that will be used repeatedly in this work.  
We will start by defining the effective  
one-particle Hamiltonian as 
\begin{align} 
        h(\mathbf{r};\mathbf{R}) 
        = \frac{-\hbar^2}{2m} \nabla^2 
        + v_\mathrm{ext}(\mathbf{r};\mathbf{R}) + v_\mathrm{eff}(\mathbf{r}). 
\end{align} 
The one-particle Hamiltonian is the starting 
point for treating electron-electron  
correlation in many-electron systems.  
In the above expression,  
$v_\mathrm{eff}$ is the  
effective one-particle operator for treating electron-electron  
interaction, which can be approximated using  
HF, DFT, pseudopotential, MBPT, or some model potential 
\begin{align} 
        v_\mathrm{eff} 
        &= 
        \{  
        v_\mathrm{HF},         v_\mathrm{KS},         v_\mathrm{ps}, v_\mathrm{model}, 
        \dots \} . 
\end{align}  
For wave function based methods,  
the eigenfunctions of the one-particle Hamiltonian  
provide the one-particle basis for  
performing, MP2, CI, and, CCSD calculations.  
\begin{align} 
        h \chi_p 
        &= 
        \epsilon_p \chi_p. 
\end{align}  
In case of DFT, these one-particle basis functions 
are used to construct the one-particle density.  
The derivation presented here is very general and does not 
make any \textit{a priori} assumption about the form of the  
effective potential.  
Typically, $v_\mathrm{eff}$ is not 
known in advance and has to be constructed  
iteratively using a self-consistent procedure. 
However, because of the SCF procedure,  
the final SCF-converged  $v_\mathrm{eff}$ 
has a functional dependence on the nuclear  
geometry $\mathbf{R}$ 
\begin{align} 
        h \chi_p 
        &= 
        \epsilon_p \chi_p 
        \xrightarrow[\textrm{convergence}]{\textrm{SCF}} 
        v_\mathrm{eff,SCF}(\mathbf{r})[\mathbf{R}]. 
\end{align} 
Using the SCF converged effective potential 
for any nuclear geometry $\mathbf{R}^\eta$, 
we define the following one-particle Hamiltonian 
\begin{align} 
        f^\eta(\mathbf{r};\mathbf{R}^\eta) 
        = \frac{-\hbar^2}{2m} \nabla^2 
        + v_\mathrm{ext}(\mathbf{r};\mathbf{R}^\eta)  
        + v_\mathrm{eff,SCF}(\mathbf{r})[\mathbf{R}^\eta] 
\end{align} 
We also define a reference structure $\mathbf{R}^0$ 
and the effective Hamiltonian corresponding to  
the reference structure is given by $f^0$: 
\begin{align} 
        f^0(\mathbf{r};\mathbf{R}^0) 
        = \frac{-\hbar^2}{2m} \nabla^2 
        + v_\mathrm{ext}(\mathbf{r};\mathbf{R}^0)  
        + v_\mathrm{eff,SCF}(\mathbf{r})[\mathbf{R}^0]. 
\end{align} 
Typically, the reference structure is selected to be the 
minimum energy structure. However, the derivation  
does not impose this as a requirement.

\subsection{Definition of deformation potential} 
\label{sec:vdef} 
We define a structure $\mathbf{R}^\eta$ 
to be deformed if is different from the  
reference structure $\mathbf{R}^0$, 
and cannot  be generated from  
the reference structure by simple rotational 
and translational transformation on $\mathbf{R}^0$. 
The difference between $\mathbf{R}^\eta$ 
and $\mathbf{R}^0$ is a consequence of  
deformation in the internal coordinates 
of the molecular system. 
Associated with each deformed structure,  
we define 
the  deformation potential, which is defined as 
\begin{align} 
\label{eq:def_vdef} 
        v^\eta_\mathrm{def} 
        &=   
        f^\eta - f^0.  
\end{align} 
Using the above definition, the deformation  
potential can be interpreted as the  
potential which must be added to  
the reference Hamiltonian to generate 
the deformed Hamiltonian.  
The existence of the  
deformation potential is an  
exact condition, and can be proved using the  
following argument. Because $f^\eta$ and  
$f^0$ exist, therefore, using Eq. \eqref{eq:def_vdef}  
$v^\eta_\mathrm{def}$ must also exist.  
 
Up until this point, we have used abstract  
Dirac representation for representing the  
reference and deformed Hamiltonians. 
However, now we seek to represent the  
deformed potential as a matrix in a single-particle basis.  
The matrix representation of the above equation  
is given as  
\begin{align} 
        \mathbf{F}^\eta = \mathbf{F}^0 + \mathbf{V}^\eta_\mathrm{def},  
\end{align} 
where the matrix elements are defined using  
as set of single-particle basis functions$\{\phi_\mu^{[\eta]} \}$ 
\begin{align} 
        F^{\eta}_{\mu \nu} 
        &= 
        \langle \phi_\mu^{[\eta]}  
        \vert          
                \mathrm{f}^\eta 
        \vert  
        \phi_\nu^{[\eta]} \rangle. 
\end{align}  
The choice of the single-particle basis functions 
is an important one, and in principle, may  
or may not depend on the nuclear geometry $\mathbf{R}^\eta$. 
The choice of the single-particle basis in which the  
deformed potential is represented is an important topic 
and will be discussed in ~\autoref{sec:implementation}.  
 
\subsection{Conceptual construction of Effective Stochastic Potential} 
\label{sec:esp} 
Calculation of the deformation potential  
becomes computationally expensive for a large set of structures.  
The effective stochastic potential method is designed to address  
this computational bottleneck by replacing the exact deformation potential 
by the effective potential that shares common identical statistical metrics with the deformation potential for a set of structures.  
Conceptually, the ESP potential can be defined by the following three steps. 
First, we define a set of all possible unique structures that a chemical system can exist in by $\{\mathbf{R}^\eta\}_0^\infty$. Second, for each of the  structures, the deformation potential is determined, where the set of all deformed potentials is defined as  
\begin{align} 
 \mathcal{S}_\mathrm{def}^V = \{\mathbf{V}^\eta_\mathrm{def}\}_0^\infty. 
\end{align} 
For each deformed structure $\mathbf{R}^\eta$, we associate a corresponding  
probability of $p^\eta$ for existence of that structure in a physical system. 
As the definition of probability implies, $p^\eta$ should satisfy the following 
two mathematical relationships: 
\begin{align} 
        p^\eta \ge 0, 
\end{align}  
and  
\begin{align} 
        \sum_{\eta \in \mathcal{S}_\mathrm{def} } 
        p^\eta  
        &= 1. 
\end{align} 
For the remainder of the derivation we will  
use the compact notation of $v_{ij}^\eta$ 
to refer to the individual matrix elements 
of the deformation potential 
\begin{align} 
        v_{ij}^\eta 
        \equiv  
        \left[ \mathbf{V}^\eta_\mathrm{def} \right]_{ij}. 
\end{align} 
For each matrix element of $\mathbf{V}^\eta_\mathrm{def}$ 
the mean $\mu$ or the expectation value of the distribution is  
defined as  
\begin{align} 
        s_{ij}^{(1)} \equiv \mu_{ij} 
        &= 
        \sum_{\eta \in \mathcal{S}_\mathrm{def} } 
        p^\eta v_{ij}^\eta. 
\label{eq:mean} 
\end{align} 
Here, we have used both $s_{ij}^{(1)}$ and $\mu_{ij}$ 
to refer to the mean for convenience.  
Analogously, the variance $\sigma$ of the distribution is defined as 
\begin{align} 
        s_{ij}^{(2)} \equiv  
        \sigma_{ij} 
        &= 
        \sum_{\eta \in \mathcal{S}_\mathrm{def} } 
        p^\eta [v_{ij}^\eta-\mu_{ij}]^2. 
\label{eq:variance} 
\end{align} 
Using the mean $\mu_{ij}$, we define the m$^\mathrm{th}$ central moment of  
the distribution by $s_m$ which is evaluated using the  
following expression: 
\begin{align} 
        s_{ij}^{(m)} 
        &= 
        \sum_{\eta \in \mathcal{S}_\mathrm{def} } 
        p^\eta [v_{ij}^\eta-\mu_{ij}]^m 
        \quad \quad \mathrm{with} \quad (m > 1). 
\end{align} 
Evaluation of all the moments for each matrix element 
of the set of deformation potentials   
implies the following relationship: 
\begin{align} 
        \{\mathbf{V}^\eta_\mathrm{def}\}_0^\infty 
        \rightarrow \{ \mathbf{s}^{(1,\mathrm{def})}, \mathbf{s}^{(2,\mathrm{def})}, \dots, \mathbf{s}^{(\infty,\mathrm{def})} \}. 
\end{align} 
Using these statistical metrics, we are now in the position  
to define the effective stochastic potential in matrix representation.  
We start by defining a symmetric stochastic matrix whose matrix elements 
are random numbers $Z$ drawn from a probability-distribution function  
$f_{ij}^\mathrm{pdf}(z)$: 
\begin{align} 
        v_{ij}^\mathrm{sto}  = v_{ji}^\mathrm{sto}  
        = Z . 
\end{align} 
The probability distribution function must satisfy the following  
two conditions: 
\begin{align} 
        f_{ij}^\mathrm{pdf}(z)  \ge 0,   \\  
        \int_{-\infty}^{+\infty} dz \quad f_{ij}^\mathrm{pdf}(z) = 1. 
\label{eq:pdf_constraints} 
\end{align} 
The probability of finding $Z$ in the interval $[a,b]$ is given by 
\begin{align} 
        P[a \le Z \le b] = \int_{a}^{b} dz \quad f_{ij}^\mathrm{pdf}(z). 
\end{align} 
We define the  
effective stochastic potential as the 
stochastic potential 
whose central-moments are  
closest to the central-moments of  
the deformed potential.  
Mathematically, the ESP potential 
is defined using the following  
constrained functional minimization: 
\begin{align} 
        \min_{f^\mathrm{pdf}_{ij}}  
        \left[ 
        \sum_{m=1}^\infty  
        \left[ s_{ij}^{(m,\mathrm{def})} - s_{ij}^{(m,\mathrm{sto})} \right]^2 
        \right] 
        \rightarrow 
        f^\mathrm{pdf,min}_{ij} 
        \rightarrow         
        v_{ij}^\mathrm{esp}[f^\mathrm{pdf,min}_{ij}], 
\end{align} 
where the constraints on $f^\mathrm{pdf}_{ij}$ 
are given in ~\autoref{eq:pdf_constraints}. 
\section{Computational and Implementation details} \label{sec:implementation}
In this section, we present the implementation details of the ESP
method by introducing additional approximations needed for practical implementation of the method. 
Practical implementation of the ESP method
requires us to work with finite sets of data. 
As a consequence, only a finite set of structures 
$\{\mathbf{R}^\eta\}_0^M$ were used for evaluation of the
deformation potential and the set of deformation potentials was 
also finite. Computational implementation also required
us to limit the set of moments $(\mathbf{s}^{(1)},\dots,\mathbf{s}^{(N)})$ on the set of deformation potentials to a finite number.
The various methods of sampling, choice of single-particle basis functions, and determination of the probability distribution function for ESP are presented below. 

\subsection{Canonical ensemble sampling using Monte Carlo and molecular dynamics}
\label{sec:md}
The probability $p^\eta$ associated with the structure $\mathbf{R}^\eta$
depends on the thermodynamic conditions of the physical system,
and can be chosen to describe both equilibrium and non-equilibrium conditions. 
In this work, we are interested in systems at thermal equilibrium, which are well-described using
the constraints of canonical ensemble with constant composition, temperature, and volume (N,V,T).
For canonical ensemble, the probability is the well-known Boltzmann expression as shown below:
\begin{align}
    p^\eta 
    &= 
    \frac{e^{-\beta (E^\eta-E^0)}}
    {\sum_{\lambda \in \mathcal{S}_\mathrm{def}} e^{-\beta (E^\lambda-E^0)}},
\label{eq:boltzmann}
\end{align}
where, $\beta = (k_B T)^{-1}$ and $E^\eta$ is the energy of the deformed structure and $E^0$ is the energy associated with 
the minimum-energy structure. 
The first step in the determination of the deformation potential is the generation 
of the set of structures $\mathbf{R}^\eta$.
In principle, the set of structures can be achieved using either Monte Carlo (MC) by molecular dynamics (MD) simulations. 
The key point in both of these approaches is to
generate a statistically meaningful sample of thermally accessible structures. 
In the MC procedure, this is achieved by 
randomly distorting the optimized structure,
calculating its energy, and 
and calculating the mean and central moments
using the Boltzmann-weighted procedure described in ~\autoref{eq:boltzmann}.
In the MD procedure, equilibrium molecular dynamics calculations
are performed at fixed N,V,T and the structures are then randomly 
selected from the equilibrium distribution. 
It is important to note that Boltzmann weights are only 
needed for MC sampling, and should not be included in the 
case of MD sampling. The relevant equations for 
MD sampling are presented in ~\autoref{eq:md_sampling}
\begin{align}
    \bm{\mu} 
    &=
    \sum_{\eta \in \mathcal{S}_\mathrm{def}^\mathrm{MD} }
     \mathbf{V}^\eta_\mathrm{def}.
\label{eq:md_sampling}
\end{align}
In the above equation, the superscript (MD) implies that
the set of structures were obtained from 
the MD simulation. 

\subsection{One-particle basis for representing the deformation potential }
\label{sec:basis}
An important feature of the ESP method is the choice of the one-particle
basis functions that are used for representing the deformation potential. 
In this work, we will use the eigenvectors of the reference Hamiltonian for representing both the deformed and the ESP potential. 
We start by defining the AO-basis representation of the 
reference Hamiltonian matrix
\begin{align}
    \mathbf{F}^\eta \mathbf{C}^\eta &=  \mathbf{S}^\eta \mathbf{C}^\eta \bm{\epsilon}^\eta.
\label{eq:basis1}
\end{align}
In the first step, we perform canonical orthogonalization 
\begin{align}
    \mathbf{X}^{\eta \dagger} \mathbf{F}^\eta \mathbf{X}^{\eta}
    &= 
    \tilde{\mathbf{F}}^\eta \\
    \mathbf{X}^{\eta \dagger} \mathbf{S}^\eta \mathbf{X}^{\eta}
    &= 
    \bm{I} .
\label{eq:basis2}
\end{align}
In the next step, we obtain the unitary matrix that diagonalizes the 
reference transformed Fock matrix $\tilde{\mathbf{F}}^0$
\begin{align}
\mathbf{U}^{0 \dagger} \tilde{\mathbf{F}}^0 \mathbf{U}^{0}
    &= 
    \tilde{\tilde{\mathbf{F}}}^0 
    \equiv
    \bm{\epsilon}^0.
\label{eq:basis3}
\end{align}
Using $\mathbf{U}^{0}$, we transformed all 
the deformed Fock matrices $\tilde{\mathbf{F}}^\eta$ into the 
eigenbasis of the reference Fock matrix $\tilde{\mathbf{F}}^0$
\begin{align}
    \mathbf{U}^{0 \dagger} \tilde{\mathbf{F}}^\eta \mathbf{U}^{0}
    &= 
    \tilde{\tilde{\mathbf{F}}}^\eta.
\label{eq:basis4}
\end{align}
It is important to note that we are transforming using $\mathbf{U}^{0} $
instead of $\mathbf{U}^{\eta}$. 
Finally, we obtain the matrix representation of the deformed potential 
in the eigenbasis of the reference Fock matrix using the following 
expression
\begin{align}
    \mathbf{V}_\mathrm{def}  = \tilde{\tilde{\mathbf{F}}}^\eta - \tilde{\tilde{\mathbf{F}}}^0,
\label{eq:basis5}
\end{align}
which is equivalent to subtracting the eigenvalues of the reference Fock matrix from the
deformed Fock matrix. 
\begin{align}
    \mathbf{V}_\mathrm{def}^\eta  = \tilde{\tilde{\mathbf{F}}}^\eta - \bm{\epsilon}^0
\label{eq:basis6}
\end{align}

\subsection{Stochastic sampling using Gaussian random matrices}
\label{sec:gauss}
In this work, we have restricted the 
sampling to only the mean and the variance
\begin{align}
    \{ \mathbf{V}_\mathrm{def}^\eta  \} 
 \rightarrow
 \{ \bm{\mu}^\mathrm{def}, \bm{\sigma}^\mathrm{def} \}.
\end{align}
The above sampling condition is computationally
implemented using symmetric Gaussian random matrices. 
Each element of the ESP matrix is 
is a Gaussian random number $Z$ as shown in ~\autoref{eq:gauss_rand_sampling}
\begin{align}
    \left[ \mathbf{V}_\mathrm{esp} \right]_{ij}
    =
    \left[ \mathbf{V}_\mathrm{esp} \right]_{ji}
    =
    v^\mathrm{esp}_{ij} 
    = Z
    \quad \mathrm{where} \quad
    Z \in \{ f_{ij}^\mathrm{Gauss}(z) \}.
\label{eq:gauss_rand_sampling}
\end{align}
The function  $f_{ij}^\mathrm{Gauss}$
is the Gaussian or normal probability distribution function 
with mean $\mu$ and variance $\sigma$, and has the following form
\begin{align}
    f_{ij}^\mathrm{Gauss}(z)
    &=
    \frac{1}{\sqrt{2 \pi {\sigma^{\mathrm{def}}_{ij}}^2 }} 
    \exp[-\frac{(z-\mu_{ij}^\mathrm{def})^2}{2 {\sigma^{\mathrm{def}}_{ij}}^2}].
\label{eq:normalsampling}
\end{align}
The corresponding Fock matrix associated with ESP is given by 
\begin{align}
    \tilde{\tilde{\mathbf{F}}}_\mathrm{esp}
    &=
    \tilde{\tilde{\mathbf{F}}}^0 + \mathbf{V}_\mathrm{esp}(Z).
\label{eq:esp_fock}
\end{align}
\section{Results}\label{sec:results}
\subsection{Statistical distribution of HOMO-LUMO gap of $\mathrm{H}_2 \mathrm{O}$ at 300 K}
We begin with a proof-of-concept calculation for the ESP method, with the goal of demonstrating that the properties of a distribution of thermally-accessible geometries of a chemical system can be accurately reproduced with a stochastic potential derived from that system. We use a single isolated water molecule as our test case. This serves as a useful example of a simple, well-studied polyatomic system. Starting with NVT ensemble, our goal is to calculate the frequency distribution of the HOMO-LUMO gap of water at 300K. 

\subsubsection{System setup and electronic structure calculations}
The calculations were performed using the canonical ensemble with constant N, V, and T at $300 K$.
The sampling of the structures for this ensemble was performed using the Monte Carlo procedure. 
A set of 1000 structures of water molecules were generated by introducing random distortions to the bond-distances and bond-angles of the optimized structure of water.  For each of the 1000 structures, Hartree-Fock (HF) calculations were performed using cc-pVDZ basis functions, 
using the GAMESS software package.~\cite{gamessref}
The probability associated with each of the 1000 structures were calculated using the 
Boltzmann expression as shown below
\begin{align}
    p_i = \frac{e^{-\frac{E_i - E_0}{k_BT}}}{\sum_j {e^{-\frac{E_j - E_0}{k_BT}}}},
\end{align}
where $E_0$ is the energy of the minimum-energy structure. 
The eigenvalues of the Fock matrix from the converged HF calculations were used to 
calculate the HOMO-LUMO gap for the 1000 structures.
\begin{align}
    \Delta \epsilon^\mathrm{HF}
    &= \epsilon_\mathrm{LUMO}^\mathrm{HF} - \epsilon_\mathrm{HOMO}^\mathrm{HF}
\end{align} 
Using the thermal probabilities $p_i$, the thermally-weighted frequency distribution of the HOMO-LUMO gaps for 1000 structures obtained from the HF/cc-pVDZ calculation at 300K is presented in ~\autoref{fig:water_gaps}. The mean and variance of the distribution is presented in ~\autoref{tab:water}. 

\subsubsection{Construction of the effective stochastic potential}
The ESP for water at 300K was constructed from a different set of 1000 Fock matrices 
obtained from the electronic structure calculations. Using ~\autoref{eq:basis1}, all the
Fock matrices were first transformed into the eigenbasis of the Fock matrix associated with the minimum-energy structure $\mathbf{F}^0$. Then, using ~\autoref{eq:basis6}, the set of 1000 deformation potential matrices were obtained. The mean and the variance associated with each matrix element of the deformation potential matrices were calculated using ~\autoref{eq:mean} and ~\autoref{eq:variance}. 
The means and variances were used in the next step to calculated the matrix elements of the 
ESP using Gaussian random numbers as show in ~\autoref{eq:normalsampling}. 
The ESP Fock matrix was constructed from the ESP matrix using ~\autoref{eq:esp_fock}. 
A set of 1000 ESP Fock matrices were generated by stochastic sampling, and the eigenvalues of the ESP Fock matrices were used to calculate the ESP HOMO-LUMO gap
\begin{align}
    \Delta \epsilon^\mathrm{ESP}
    &= \epsilon_\mathrm{LUMO}^\mathrm{ESP} - \epsilon_\mathrm{HOMO}^\mathrm{ESP}.
\end{align}

\subsubsection{Comparison of the ESP and electronic structure results}
The comparison of the mean, the variance and higher central moments of the frequency distributions of HOMO-LUMO gap energies obtained using both Hartree Fock and ESP methods are presented in ~\autoref{tab:water}. 
\begin{table}[]
\centering
\caption{Mean ($\mu$), standard deviation, ($\sigma_2$) and the $p^\mathrm{th}$ root of central moments ($\sigma_4$) of HOMO-LUMO gap energy in eV of water for Hartree Fock (cc-pVDZ basis) and ESP method at a given number of sampling points $N_\mathrm{sample}$. Water geometries were generated from a Monte Carlo procedure.}
\label{tab:water}
\begin{tabular}{llccc}
\hline
Method & $N_\mathrm{sample}$ & $\mu$    & $\sigma_2$          & $\sigma_4$     \\
\hline
HF     	&	 $1 \times 10^3$  &	14.85	& 0.0118 &	0.0155	\\
ESP    	&	 $1 \times 10^3$  &	14.85	& 0.1023 &	0.0137	\\
ESP    	&	 $5 \times 10^3$  &	14.85	& 0.1043 &	0.0146	\\
ESP    	&	 $1 \times 10^4$  &	14.85	& 0.1043 &	0.0145	\\
ESP    	&	 $5 \times 10^4$  &	14.85	& 0.1039 &	0.0142	\\
ESP    	&	 $1 \times 10^5$  &	14.85	& 0.1040 &	0.0142	\\
\hline
\end{tabular}
\end{table}
Using the mean $\mu$, the $p^\mathrm{th}$ root of the central moment is defined as 
\begin{align}
\label{eq:sigma_p}
	\sigma_p 
	&=
	\left[
	\frac{1}{N_\mathrm{sample}} 
	\sum_{i=1}^{N_\mathrm{sample}} 
	(x_i - \mu)^p
	\right]^{1/p}
\end{align}
It was found that both the mean HOMO-LUMO gap 
obtained from ESP were in good agreement with the HF results. The results demonstrate that the ESP
can successfully capture the
fluctuations present in the deformation potential at non-zero temperatures.
However, comparison of 
higher-order moments (\autoref{tab:water}) reveal that the
two distributions have similar but not identical 
characteristics. 
In order to compare the statistics of the Hartree Fock and ESP method distributions, we used Z-score (also called standard score)~\cite{urdan2011statistics}. Z-score is defined mathematically as
\begin{align}
\mathrm{Z-score} = \frac{x - \mu}{\sigma_2},
\end{align}
where $x$ is a measured value, $\mu$ is the mean of the value over a distribution, and $\sigma_2$ is the corresponding standard deviation. Z-score is a unitless quantity that is relative to the distribution's standard deviation. A positive Z-score reflects a value greater than the mean, and negative Z-score reflects a value smaller than the mean. A Z-score of 1 corresponds to a value that is exactly one standard deviation greater than the mean of the distribution.  
\begin{figure}[!ht]
\vspace{-0 pt}
\centerline{\includegraphics[width=6.00in]{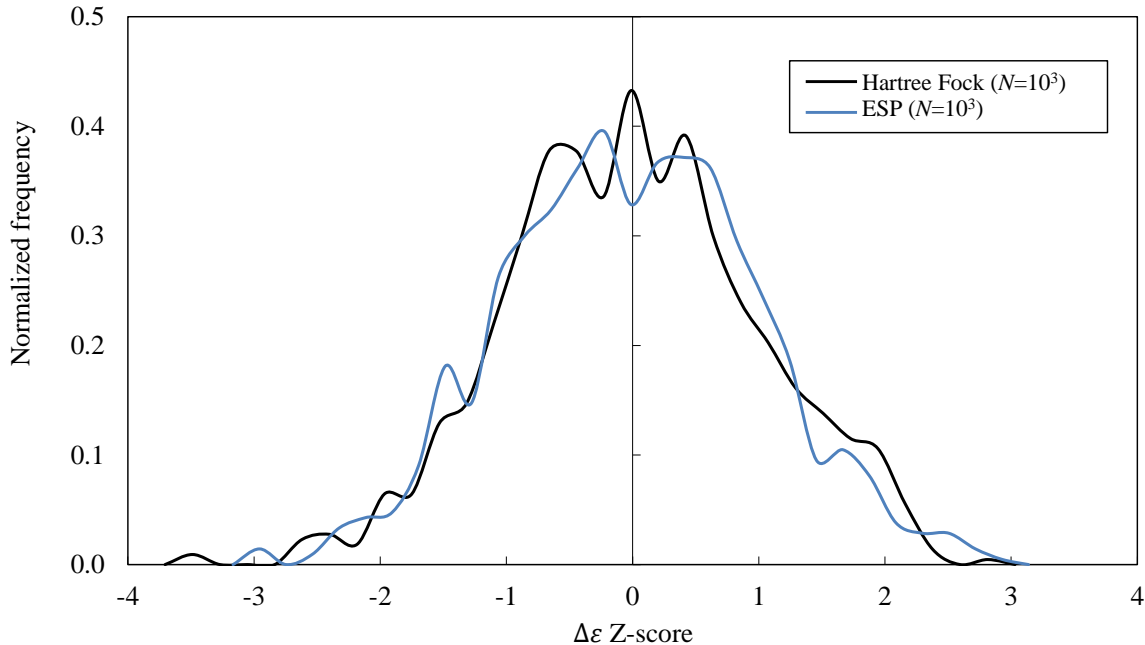}}
\vspace{-0 pt}
\caption{Comparison of HOMO-LUMO gap energy Z-Scores of a single gas-phase water molecule between Hartree Fock (cc-pVDZ basis) and ESP method.}
\label{fig:water_gaps}
\end{figure}
Although, the ESP and HF distribution were found to differ in certain regions (\autoref{fig:water_gaps}), 
the overall features of the both the distributions were found to in good agreement with 
each other. 
\subsubsection{Effect of sample size on frequency distribution}
\begin{figure}[!ht]
\vspace{-0 pt}
\centerline{\includegraphics[width=6.00in]{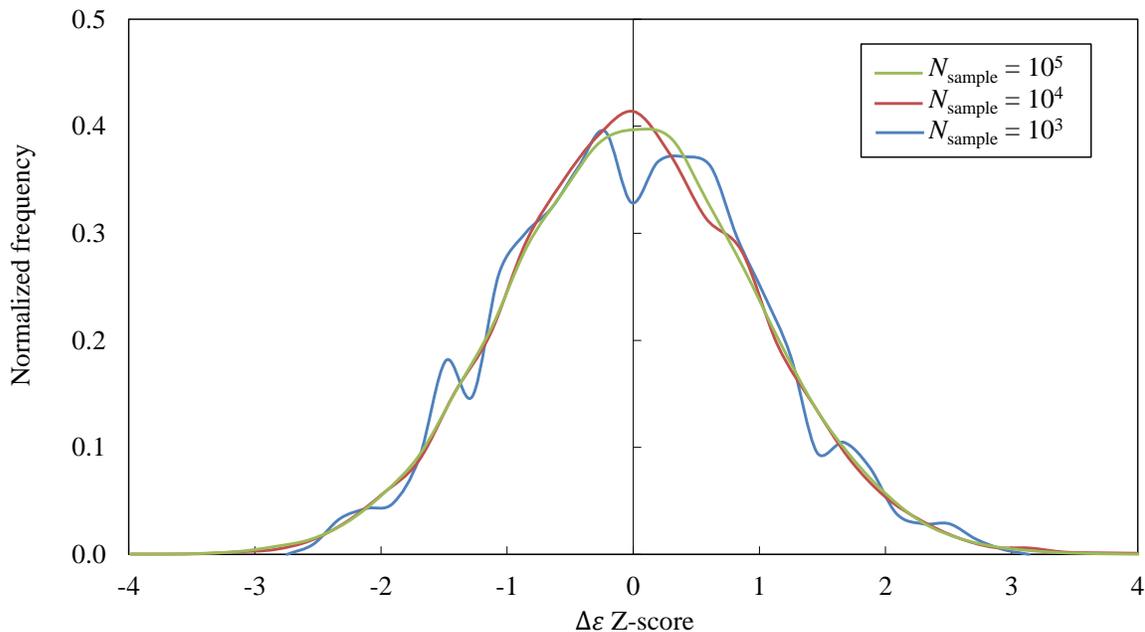}}
\vspace{-0 pt}
\caption{Comparison of HOMO-LUMO gap energy Z-Scores of gas-phase water molecule between Hartree Fock (cc-pVDZ basis) and ESP method using increasing number of sampling points.}
\label{fig:water_sample}
\end{figure}

The effect of sample size on the ESP results were investigated 
by performing  $10^3$, $10^4$, and $10^5$
uncorrelated sets of ESP calculations, and the results are summarized in 
\autoref{tab:water} and \autoref{fig:water_sample} . In all cases, the mean and higher-order moments 
exhibited convergence with the size of the sample.
The frequency distributions in \autoref{fig:water_sample} show that the 
noise in the distribution for $N_\mathrm{sample}=1000$ is attenuated 
by increasing the sampling size to $10^4$. 

\subsection{HOMO-LUMO gap of $\mathrm{Cd}_{20} \mathrm{Se}_{19}$ in water at 300K }
We now turn the discussion to the application of the ESP method to a small CdSe QD (1.2 nm in diameter) in aqueous media at 300K. These systems are excellent candidates on which to apply the ESP method, due to the potentially useful applications of QDs, the complexity of their electronic states, and the impact of solvent environment and non-zero temperature of its optical and electronic properties. 

\subsubsection{System setup and construction of the effective stochastic potential}
Equilibrium molecular dyanamics simulations with constant N,V,T 
were performed for the $\mathrm{Cd}_{20} \mathrm{Se}_{19}$ 
cluster with explicit water molecules at 300K.
The force-field developed by Rabani et al.~\cite{doi:10.1063/1.1424321}
was used for the CdSe cluster, and the 
explicit water molecules were treated using TIP3 force.
The MD calculations and analysis of the simulations were
performed by the GROMACS molecular dynamics program. 
The initial structure for the MD calculations 
was obtained from the minimum-energy structure
obtained from electronic structure theory at HF/LANL2Z ECP level. 
The initial structure was then re-optimized using the CdSe force-field 
in the MD program in vacuum. 
The simulation cell was constructed by solvating the CdSe cluster 
in a water bath consisting of 11763 water molecules. 
The entire system was equilibrated at 300K by performing MD
simulation for 500 ns. 
The equilibration conditions were verified by
monitoring the total energy, total potential energy, and
the temperature of the system.
\par
 After equilibration, a set of 50 structures was randomly selected from the MD trajectories
for construction of the ESP matrix. As a first step, HF calculation using LANL2Z ECP basis functions were 
performed for the set of structures, and the deformation potential matrices were calculated.
The mean and variance of each matrix element were calculated, and the ESP matrix and the ESP Fock matrix were constructed using the procedure described in ~\autoref{sec:implementation}. 

\subsubsection{Predictions from the ESP calculations}
The average HOMO-LUMO gap $(\mu)$ and the higher-order moments (\autoref{eq:sigma_p}) obtained using the ESP are presented in \autoref{tab:cdse}. 
\begin{table}[]
\centering
\caption{Mean ($\mu$), standard deviation, ($\sigma_2$) and the $p^\mathrm{th}$ root of central moments ($\sigma_4$) of HOMO-LUMO gap energy in eV  of Cd$_{20}$se$_{19}$ for Hartree Fock (LAN2LZ-DZ ECP basis) and ESP method at a given number of sampling points $N_\mathrm{sample}$. Cd$_{20}$se$_{19}$ geometries were generated from molecular dynamics simulation.}
\label{tab:cdse}
\begin{tabular}{llccc}
\hline
Method & $N_\mathrm{sample}$ & $\mu$    & $\sigma_2$          & $\sigma_4$     \\
\hline
HF     	&	 $5 \times 10^1$  &	2.0571 &	0.2530 &	0.0828 \\
ESP    	&	 $5 \times 10^1$  &	1.9514 &	0.5745 &	0.8697 \\
ESP    	&	 $1 \times 10^3$  &	0.9639 &	0.8087 &	1.6318 \\
ESP    	&	 $5 \times 10^3$  &	0.9806 &	0.7429 &	1.2373 \\
ESP    	&	 $1 \times 10^4$  &	0.9775 &	0.7431 &	1.2458 \\
ESP    	&	 $5 \times 10^4$  &	0.9748 &	0.7419 &	1.2719 \\
ESP    	&	 $1 \times 10^5$  &	0.9760 &	0.7446 &	1.2736 \\
\hline
\end{tabular}
\end{table}
The results from \autoref{tab:cdse} 
highlight  the 
need for large sampling size
to obtain statically meaningful quantities. 
Specifically, we find that
a sample size of $N_\mathrm{sample}=50$
is inadequate for describing the 
distribution of the HOMO-LUMO gap
for both HF and ESP calculations. 
The results from ESP show that 
sample size in excess of $1000$
are needed to converge the mean and 
higher-order moments. 
Comparison between the HF and ESP 
results show that mean HOMO-LUMO
gap is significantly over-estimated in 
HF set of calculations because of inadequate sample size. 
\begin{figure}[!ht]
\vspace{-0 pt}
%\centerline{\includegraphics[width=8.00in]{./plots/cd20se19_zscores.pdf}}
\centerline{\includegraphics[width=6.00in]{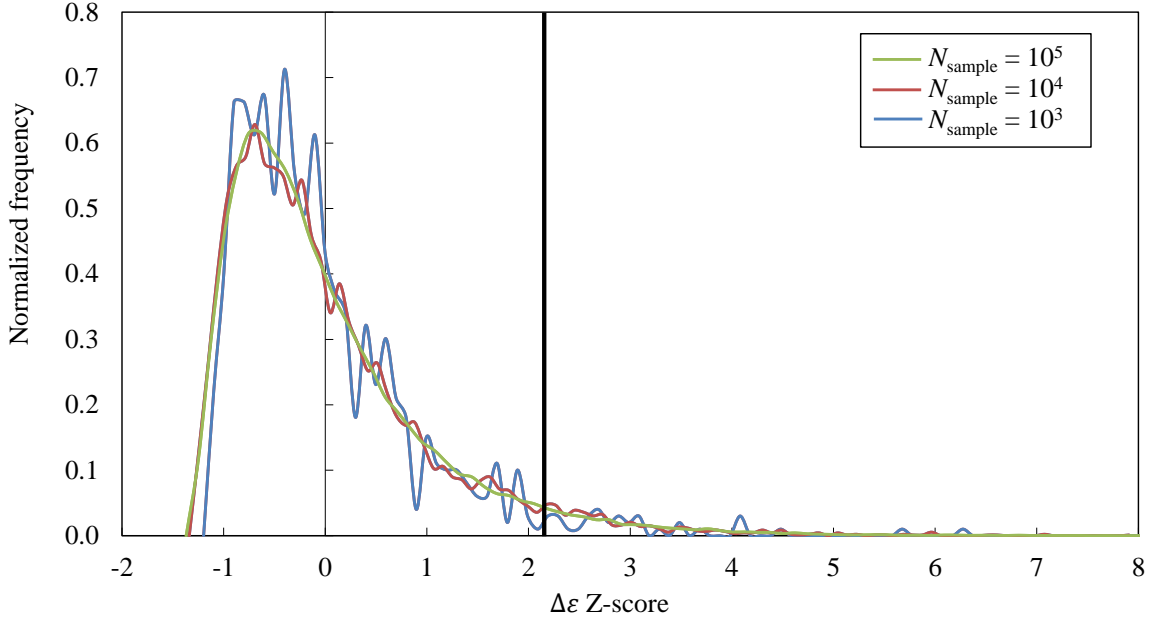}}
\vspace{-0 pt}
\caption{HOMO-LUMO gap Z-scores for a Cd$_{20}$Se$_{19}$ quantum dot using increasing number of sampling points at $T=300$K. The vertical bar at Z=2.2 represents the HOMO-LUMO gap energy of the 0K optimized structure with $\Delta \epsilon = 2.7 \mathrm{eV} $.~\cite{zungercdse}}.
\label{fig:Cd20Se19_zscore_gap}
\end{figure}
The distribution of the HOMO-LUMO gap from the ESP calculations are 
presented in \autoref{fig:Cd20Se19_zscore_gap}.
The distribution was found to be highly asymmetric with respect to the 
origin with a significant positive bias towards the -Z direction 
and a long tail in the +Z direction. 
The frequency distribution from the ESP data (\autoref{fig:Cd20Se19_zscore_gap}) implies that at 300K, 
the HOMO-LUMO gap for majority of the structures are
lower than the mean gap. The long tail implies, that 
a very small fraction of the population exhibits very 
large HOMO-LUMO gaps. 
The distribution at 300K is compared with the
the HOMO-LUMO gap of the minimum-energy structure (vertical black line in \autoref{fig:Cd20Se19_zscore_gap}). The comparison show 
that the maxima of the frequency distribution is significantly different from the 
minimum-energy structure. This implies that the minimum-energy structure is a poor representative descriptor for the set of thermally accessible structures at 300K.  From  \autoref{fig:Cd20Se19_zscore_gap}
we conclude that the heating the system from 0K to 300K 
should decrease the HOMO-LUMO gap of CdSe clusters. 
This observation from the ESP calculation also is supported by 
previous experimental~\cite{doi:10.1063/1.2357856,Savchenko:17} 
and theoretical~\cite{olegcdse,doi:10.1063/1.3526297} 
studies on CdSe  and other semiconductor clusters and quantum dots.  
\section{Discussion}\label{sec:discuss}
\textbf{Sampling constant temperature ensembles}:
This work demonstrates that the ESP method can be used with either Monte Carlo or molecular dynamics simulations. Specifically, the water calculations were 
performed using Monte Carlo sampling and the 
 $\mathrm{CdSe}$ cluster calculations were performed using classical molecular dynamics simulation. 
 In our experience, classical MD simulation provides a
 an efficient computational route for sampling NVT and NPT
 ensembles for solvated molecular systems which explicit
 solvent molecules.  We believe that the study of $\mathrm{CdSe}$ cluster
 in water are particularly relevant to the study 
of semiconductor nanoparticles, because it 
has generated new insight into the electronic structure 
of clusters at non-zero temperatures in aqueous media. 
The results from the ESP calculations also highlighted
the requirement of a large sample set for obtaining 
statistically meaningful results. 

\textbf{Future directions:}
Although the proof-of-concept calculations presented here 
were restricted to HF calculations, the ESP method 
is can be combined other methods for treating electron-electron 
correlation. 
The ESP method generates a 
manifold of single-particle states which can be
used as a starting point for combining ESP with existing many-body theories such as DFT, TDDFT, and Green's function methods. 
In addition, the ESP method can also be applied to 
non-equilibrium or non-Boltzmann distributions by appropriate 
choice of the set of probabilities $\{p^\eta\}$ (either discrete or continuous)
that govern the population of molecular structures for a given chemical environment.
\section{Conclusions}\label{sec:conclusions}
This work presented the development of the effective stochastic potential method (ESP) as a
 route to perform efficient conformational sampling for molecular systems at non-zero temperatures.
The central idea of the ESP method is to replace the fluctuations experienced by the molecular system by a stochastic potential that provides an equivalent statistical description of those fluctuations.  
The concept of deformation potential was introduced and the existence theorem for
such a potential was presented. 
The formal mathematical constructions of the ESP potential was
presented using a functional minimization approach
to capture the fluctuations in the deformation potential. 
The computational implementation of the ESP
was achieved using random matrix theory. 
Two proof-of-concept calculations were performed using the developed ESP method,
by calculating the distributions of HOMO-LUMO gaps
of water and solvated CdSe cluster at 300 K. 
The results from these calculations demonstrate the efficacy of the ESP method in providing an efficient conformational sampling of molecular systems at non-zero temperatures. 
\begin{acknowledgements}
This material is based upon work supported by the National Science Foundation under Grant No. CHE-1349892. This work used the Extreme Science and Engineering Discovery Environment (XSEDE), which is supported by National Science Foundation grant number ACI-1053575. We are also grateful to Syracuse University for additional computational and financial support. 
\end{acknowledgements}

%\bibliography{ref_qd_rmt,ref_qd_xchf,ref_rmt,ref_organic_thermal}
\providecommand{\latin}[1]{#1}
\providecommand*\mcitethebibliography{\thebibliography}
\csname @ifundefined\endcsname{endmcitethebibliography}
  {\let\endmcitethebibliography\endthebibliography}{}

\end{document}